\documentstyle[12pt]{article} 
\begin{document}

\def\be{\begin{equation}}
\def\ee{\end{equation}}


\begin{center}
{\Large\bf  Bulk antisymmetric tensor fields in a Randall-Sundrum model}\\[20mm]
Biswarup Mukhopadhyaya\footnote{E-mail: biswarup@mri.ernet.in}\\
{\em Regional Centre for Accelerator-based Particle Physics\\
Harish-Chandra Research Institute\\
Chhatnag Road, Jhusi, Allahabad - 211 019, India} 

Somasri Sen \footnote{Also at Department of Physics, St. Stephen's College, New Delhi - 110007, India,\\ E-mail: somasri.ctp@jmi.ac.in}\\
{\em Centre for Theoretical Physics\\
Jamia Milia University\\
New Delhi 110 025, India}

Soumitra SenGupta \footnote{E-mail: tpssg@iacs.res.in} \\
{\em Department of Theoretical Physics\\
 Indian Association for the 
Cultivation of Science\\
Calcutta - 700 032, India}\\
\vspace{0.1cm}
\end{center}

\begin{abstract}
We consider bulk antisymmetric tensor fields of various
ranks in a Randall-Sundrum scenario. We show that,
rank-2 onwards, the zero-modes of the projections of these
fields on the (3+1) dimensional visible brane become
increasingly weaker as the rank of the tensor increases.
All such tensor fields of rank 4 or more are absent from
the dynamics in four dimensions.
This leaves only the zero-mode graviton to have coupling
$\sim 1/M_P$ with matter, thus explaining why the large-scale 
behaviour of the universe is governed by gravity only. We have 
also computed the masses of the heavier modes upto  rank-3, 
and shown that they are relatively less likely to have detectable 
accelerator signals.
\end{abstract}

\vskip 1 true cm

\newpage
Extra spacelike dimensions with warped geometry have been shown to provide a
rather elegant solution to the problem of hierarchy between the Planck scale
and the electroweak scale, an embarrassment of the standard electroweak 
model. The exponential `warp factor' attached to the Minkowski part of the metric
associates a suppression factor for all masses and (gravitational) couplings of 
known fields residing on the `visible brane' localised at one of the 
orbifold fixed points of the extra compact spacelike dimension. This is the
essence of a Randall-Sundrum (RS) type of theory \cite{rs}.

While the standard model (SM) fields are assumed to lie on a 3-brane in this 
scenario, the essential input is that gravity propagates in a 5-dimensional
anti-de Sitter bulk spacetime \cite{brane1} . A natural explanation of such a description comes
from string theory, with the SM fields arising as excitation modes of an open string
whose ends lie on the brane. The graviton, on the other hand, is a closed string excitation
and hence its non-localisation seems to be in order \cite{string}. Consequently, when one takes its
projections on the visible brane, massless
graviton mode has a coupling $\sim 1/M_P$ with all matter, 
while the massive modes have enhanced coupling through the warp factor. It not only 
accounts for the observed impact of gravity in our universe but also raises hopes for
new signals in accelerator experiments \cite{rssig}.  However, there are various antisymmetric 
tensor fields which also comprise excitations of a closed string \cite{string}, 
and therefore can be expected to lie in the bulk similarly as gravity. The question
we ask here is: can these fields also have observable effects? If not, why are
the effects of their massless modes less perceptible than the force of gravitation?

Bulk fields other than gravitons have been studied earlier in RS scenarios, starting
from bulk scalars which have been claimed to be required for 
stabilisation of the modulus \cite{gw}.
Bulk gauge fields and fermions have been considered, too, with various phenomenological
implications \cite{davod}. While some of such scenarios are testable 
in accelerator experiments \cite{bulkac}  or observations in the neutrino sector \cite{gross}, 
by and large they do not cause any contradiction
with our observations so far. 

However, the situation with tensor fields of various ranks (higher than 1) is slightly 
different. For example, as has been already noted, 
an antisymmetric rank-2 tensor field such as the Kalb-Ramond 
excitation \cite{kr} can be in the bulk as 
legitimately as the graviton, and {\it prima facie} 
has similar coupling to matter as gravity. Using a generalised form of the 
Einstein-Cartan action, it has been shown that such a field is equivalent 
to torsion in spacetime \cite{ssgpm}, on which the experimental limits are 
quite severe \cite{torlim}. This apparent contradiction has been ameliorated 
in an earlier work \cite{ourprl} where it has been shown 
that the zero mode of the antisymmetric tensor field gets an additional exponential
suppression compared to the graviton on the visible brane. This could well be an
explanation of why we see the effect of curvature but not of torsion in the
evolution of the universe. Arguments in this line will however be complete only when
we can similarly address the effects of other, higher rank, antisymmetric fields which 
occur in the  NS-NS or RR sector of closed string excitations \cite{ gsw}. We address that 
question in the current study, in the special context of RS-like models. What we wish to 
point out as a whole is that the zero mode
of {\em any antisymmetric tensor field} undergoes  progressive exponential suppression 
increasing with the rank of the tensor. Moreover, for most  higher rank tensor fields it becomes
impossible to have non-vanishing components on the brane, partly because of the antisymmetric
nature of the tensor, and partly due to the gauge freedom of these fields, which reduces
the available degrees of freedom on the brane to zero.

In order that a rank-n antisymmetric tensor gauge field $X_{a{_1}a_{2}...a{n}}$ can be part 
of the dynamics, one should be able to write down a rank-(n+1) field strength tensor 

\begin{equation}
Y_{a{_1}a_{2}...a{n+1}} = \partial_{[a_{n+1}}X_{a{_1}a_{2}...a{n}]}
\end{equation}

Since a spacetime of dimension D admits of a maximum rank D for an antisymmetric tensor,
one can at most have $(n+1) = D$. Thus any antisymmetric tensor field X can have
a maximum rank $D-1$, beyond which it will all have either zero components or will become
an auxiliary field with the field strength tensor vanishing identically.
Such an auxiliary  field can be eliminated via the equations of motion if it has no mass term 
in the bulk, a feature shared by all antisymmetric tensor excitations of a closed string due
to gauge invariance.

Now let us consider a $3$-brane in an RS-type 5-dimensional 
anti-de Sitter bulk spacetime, 
where the extra spatial dimension has been compactified on an 
$S_1 /Z_2$ orbifold. There are two branes
at the orbifold fixed points $\phi = 0$ and $\pi$, where $\phi$ is the
angular variable corresponding to the compact dimension. In such a scenario, 
the 5-dimensional metric can be written as 

\begin{equation}
ds^2=e^{-2\sigma}\eta_{\mu\nu}dx^{\mu}dx^{\nu}+r_c^2d\phi^2
\end{equation}

\noindent 
with $\eta_{\mu\nu}~=~(-,+,+,+)$, and $\sigma~=~k{r_c}|\phi|$. 
$r_c$ is the radius of the compact dimension $y$,
with $y = r_c \phi$. $k$ is on the order of the
5-dimensional Planck mass $M$. The standard model fields reside at
$\phi~=~\pi$ while gravity peaks at $\phi~=~0$. The dimensional parameters 
defined above are related to the  4-dimensional Planck scale $M_P$
through the relation

\begin{equation}
M_P^2=\frac{M^3}{k}[1-e^{-2kr_c\pi}]
\end{equation}

Clearly, $M_P$, $M$ and $k$ are all of the same order of magnitude. 
For $k r_c~\simeq~12$ the exponential factor (frequently referred to as the
`warp factor') produces TeV scale mass parameters 
(of the form $m~=~Me^{-kr_{_c}\pi}$) 
on the visible brane. Thus the hierarchy between the Planck 
and TeV scales is achieved without fine-tuning.

The closed string modes of excitation pertinent to such a scenario
are antisymmetric tensor fields of various ranks, 
in addition to the graviton. Following the reasoning
given above, such fields can at most be of rank-4. A rank-5 field
has a rank-6 field strength tensor in the kinetic energy term, which, 
by virtue  of its complete antisymmetry, cannot exist in 5-dimensions.
Thus such a field can be removed using the equations of motion, while 
fields of even higher rank themselves vanish identically.

Taking a closer look at the rank-5 field strength tensor $Y_{ABCMN}$
of a rank-4 field, one gets two kinds of terms,namely :

\begin{equation}
Y_{ABCMN} = \partial_{[\mu}X_{\nu\alpha\beta y]}
\end{equation}
and
\begin{equation}
Y_{ABCMN} = \partial_{[y}X_{\mu\nu\alpha\beta]} 
\end{equation}

\noindent
where the Latin indices denote bulk co-ordinates, the Greek indices
run over the (3+1) Minkowski co-ordinates and $y$ stands for the compact
dimension. The first class of terms can be removed using the gauge freedom

\begin{equation}
\delta X_{ABCM} = \partial_{[A} \Lambda_{BCM]}
\end{equation}

\noindent 
which allows the use of 10 gauge-fixing conditions for an
antisymmetric $\Lambda_{BCM}$. As a result one can use 

\begin{equation}
X_{\nu\alpha\beta y} = 0
\end{equation}

The second class of terms do not yield any kinetic energy for
$X_{\mu\nu\alpha\beta}$ on the visible brane, and they can 
thus be removed using the equation of motion.\footnote{In principle,
such an auxiliary field can have an interaction term of the
form $X_{\mu\nu\alpha\beta}B^{\mu\nu}B^{\alpha\beta}$ with second
rank antisymmetric tensor fields. If such terms at all exist,
they will at most result in quartic self-couplings of the
rank-2 field}. Thus the rank-4 antisymmetric tensor fields (and
of course those of all higher ranks) have no role to play in the
four-dimensional world in the RS scenario.

Thus all that can matter are the lower rank antisymmetric tensor fields.
The case of a rank-2 field in the bulk (known as the Kalb-Ramond field) has
been already investigated, leading to the rather interesting observation that
the warped geometry results in an additional exponential suppression of its
zero mode on the visible brane with respect to the graviton. This suggests
an explanation of why torsion can be imperceptible relative to curvature
in our four-dimensional universe.

Let us now consider the only antisymmetric tensor field of 
a higher rank, surviving on the 3-brane. This is a rank three
tensor $X_{MNA}$, with the corresponding field strength
$Y_{MNAB}$. The action for such a field in 5-dimensions is

\be
S=\int d^5 x\sqrt{-G} Y_{MNAB}Y^{MNAB}
\ee
where $G$ is the determinant of the 5-dimensional metric.
Using the explicit form of the RS metric and taking into 
account the gauge fixing condition $X_{\mu\nu y}=0$, 
one obtains

\be
S_x=\int{d^4 x\int{d\phi[e^{4\sigma}\eta^{\mu\lambda}\eta^{\nu\rho}\eta^{\alpha\gamma}
\eta^{\beta\delta}Y_{\mu\nu\alpha\beta}Y_{\lambda\rho\gamma\delta}+
4 \frac{e^{2\sigma}}{r_c^2}\eta^{\mu\lambda}\eta^{\nu\rho}
\eta^{\alpha\gamma}\partial_{\phi}X_{\mu\nu\alpha}\partial_{\phi}X_{\lambda\rho\gamma} ]}}
\ee

Considering the Kaluza Klein decomposition of the field X,

\be
X_{\mu\nu\alpha}(x,\phi)=\Sigma_{n=0}^{\infty}X^n_{\mu\nu\alpha}(x)\frac{\chi^n(\phi)}{\sqrt{r_c}}
\ee
an effective action of the following form can be obtained in terms of the 
projections  $X^n_{\mu\nu\alpha}$ on the visible brane:

\be
S_X=\int{d^4 x\Sigma_n[\eta^{\mu\lambda}\eta^{\nu\rho}\eta^{\alpha\gamma}\eta^{\beta\delta}
Y^n_{\mu\nu\alpha\beta}Y^n_{\lambda\rho\gamma\delta}+ 4 m_n^2 \eta^{\mu\lambda}\eta^{\nu\rho}
\eta^{\alpha\gamma}X^n_{\mu\nu\alpha}X^n _{\lambda\rho\gamma}]}
\ee
where $m_n^2$ is defined through the relation
\be
-\frac{1}{r_c^2}\frac{d}{d\phi}(e^{2\sigma}\frac{d}{d\phi}{\chi^n})=m_n^2\chi^n e^{4\sigma}
\ee
and $\chi^n$ satisfies the orthonormality condition
\be
\int{e^{4\sigma}\chi^m(\phi)\chi^n(\phi) d\phi}=\delta_{mn}
\ee

In terms of $z_n=\frac{m_n}{k}e^{\sigma}$ and $f_n=e^{\sigma}\chi^n$, equation (12)
can be recast in the form 
\be
[z_n^2 \frac{d^2}{dz_n^2}+z_n\frac{d}{dz_n}+(z_n^2-1)]f_n=0
\ee
which is a Bessel Equation of order 1. 

The solution for $\chi^n$ is given by
\be
\chi^n=e^{-\sigma}f_n=\frac{e^{-\sigma}}{N_n}[J_1(z_n)+\alpha_n Y_1(z_n)]
\ee 
where $J_1(z_n)$ and $Y_1(z_n)$ respectively are Bessel and Neumann 
functions of order 1. $\alpha_n$ and $N_n$ are integration constants which can 
be determined from orthogonality and the continuity conditions at the orbifold fixed 
points. In addition, the continuity condition for the derivative 
of $\chi_n$ at $\phi=0$ yields 
\be
\alpha_n=-\frac{J_2(\frac{m_n}{k})}{Y_2(\frac{m_n}{k})}
\ee

Using the fact that $e^{kr_c\pi}>>1$ and the mass values $m_n$ on the brane is expected to 
be on the order of TeV scale $(<<k)$,
\be
\alpha_n\sim \frac{\pi}{2^5}(\frac{m_n}{k})^4<<1
\ee

The boundary condition at $\phi=\pi$ gives
\be
J_2(x_n)=0
\ee 
where $x_n=z_n(\pi)= \frac{m_n}{k}e^{kr_c\pi}$.
The roots of the above equation determine 
the masses of the higher excitation modes. As $x_n$ is 
of order unity, the massive modes lie in the TeV scale. 

Furthermore, the normalisation condition yields
\be
N_n=\frac{e^{kr_c\pi}}{\sqrt{kr_c}}J_1(x_n)
\ee
and the massive modes can be obtained from the equation

\be
\chi^n(z_n)=\sqrt{kr_c}\frac{e^\sigma}{e^{kr_c\pi}}\frac{J_1(z_n)}{J_1(x_n)}
\ee

The values of the first few massive modes of the rank-3 antisymmetric tensor field
are listed in Table 1, where we have also shown the masses of the graviton 
as well as the rank-2 antisymmetric Kaluza-Klein modes. It can be noticed that
the rank-3 field has higher mass than the remaining two at every order, and, while
the Kalb-Ramond massive modes  can have some signature at, say,
the Large hadron collider (LHC), that of the rank-3 massive tensor field
is likely to be more elusive.

\begin{center}
$$
\begin{array}{|c|c c c c|}
\hline
n  & 1  & 2  & 3  & 4  \\
\hline
m_n^{grav}~(TeV) & 1.66 & 3.04 & 4.40 & 5.77 \\
\hline
m_n^{KR}~(TeV)  & 2.87 & 5.26 & 7.62 & 9.99 \\
\hline
m_n^{X}~(TeV)  & 4.44 & 7.28 & 10.05 & 12.79 \\
\hline
\end{array}
$$ 
\end{center} 


{\em Table 1: The masses of a few low-lying modes of the graviton,
   Kalb-Ramond (KR) and rank-3 antisymmetric tensor (X) fields, 
   for $kr_c=12$ and $k=10^{19}$Gev.}

\vspace{0.2cm}

Finally, and most crucially, we examine the massless mode,
whose strength on the brane needs to be compared to that of the graviton
and the rank-2 field. The solution for this mode is given by
\be
\chi_0=-\frac{C_1}{2kr_c}e^{-2\sigma}+C_2
\ee

Requiring the continuity of $\frac{d\chi^0}{d\phi}$ at 
$\phi = \pi$, one obtains $C_1=0$. The 
normalisation condition finally gives
\be
\chi^0=\sqrt{2kr_c}e^{-2kr_c\pi}
\ee

This shows that the zero mode of the rank-3 antisymmetric tensor field
is suppressed by an additional exponential factor relative to the 
corresponding rank-2 field which already has an exponential suppression
compared to the zero mode of the graviton. Using the same argument as in
reference \cite{ourprl}, one can translate this result into the coupling of
the field X to matter, and show that the interaction with, say,
spin-1/2 fields is suppressed by a factor $e^{-2kr_c\pi}$.
Thus the higher order antisymmetric
field excitations have progressively insignificant roles to play on the
visible brane, with the fields vanishing identically beyond rank 3.

In conclusion, the graviton seems to have a unique role among the
various closed string excitations in a warped geometry. This is because
the intensity of its zero mode on the 3-brane leads to coupling
$\sim 1/M_P$ with matter fields, which is consistent with the part
played by gravity (or more precisely the curvature of spacetime) 
observed in our universe. On the contrary, while bulk antisymmetric
tensor fields upto rank-3 can still have non-vanishing zero modes
in four-dimensional spacetime, their  strength is progressively
diminished for ranks-2 and 3. This may well serve as an explanation of 
why the evolution of our universe is solely controlled by gravitation.
In addition, the masses of the higher modes also 
tend to increase with rank, making them less and less relevant
to accelerator experiments. 

{\bf Acknowledgement:} The work of BM was supported by funding from the
Department of Atomic Energy, Government of India, through the XIth 
Five Years Plan. SS and SSG acknowledge the hospitality of the Regional Centre 
for Accelerator-based Particle Physics, Harish-Chandra Research Institute.


\end{document}